\documentclass[english,superscriptaddress,nofootinbib,preprint]{revtex4}
\usepackage[T1]{fontenc}
\usepackage[latin1]{inputenc}
\usepackage{graphicx}
\usepackage{amssymb}

\makeatletter

\providecommand{\tabularnewline}{\\}

\makeatletter



\makeatother

\usepackage{babel}
\makeatother
\begin{document}
\newcommand{\pslash}[1]{\not{\!#1}}

\begin{flushright}\par\end{flushright}

\title{Anomalous $gt\bar{t}$ couplings in the Littlest Higgs Model with
T-parity}

\author{Qing-Hong Cao}

\email{qcao@ucr.edu}

\affiliation{Department of Physics and Astronomy, University of California at
Riverside, Riverside, CA 92521 USA}

\author{Chuan-Ren Chen}

\email{crchen@pa.msu.edu}

\affiliation{Department of Physics and Astronomy, Michigan State University, East
Lansing, MI 48824 USA}

\author{F. Larios}

\email{larios@mda.cinvestav.mx}

\affiliation{Departamento de F\'{\i}sica Aplicada, CINVESTAV-Mérida, A.P. 73,
97310 Mérida, Yucatán, México}

\author{C.-P. Yuan}

\email{yuan@pa.msu.edu}

\affiliation{Department of Physics and Astronomy, Michigan State University, East
Lansing, MI 48824 USA}

\begin{abstract}
In this work we calculate the \emph{leading} electroweak (EW) corrections
to the anomalous $gt\bar{t}$ coupling in the Littlest Higgs model
with T-parity (LHT), by applying the Goldstone boson equivalence theorem.
In the LHT model, such electroweak corrections arise from the loop
diagrams of heavy fermions and the {}``would-be'' Goldstone bosons.
We further examine the EW corrections in the top quark pair production
via the quark annihilation process at the LHC. The negative EW corrections
in the Standard Model are partially canceled by the positive EW corrections
from the loops of the new heavy particles, and the latter dominates
in the large invariant mass of the top quark pair. 
\end{abstract}

\maketitle
%\date{\today}

\section{introduction}

The top quark is a special quark in the Standard Model (SM) due to
its large mass. As the top quark mass ($m_{t}$) is close to the electroweak
symmetry breaking (EWSB) scale, $m_{t}\sim170.9$ GeV~\citep{tmass:2007bxa},
studying the top quark physics might shed lights on the mechanism
of EWSB. At the Tevatron, the top quark pair is mainly produced via
the quark-antiquark annihilation, whereas at the CERN Large Hadron
Collider (LHC) it is produced mainly through gluon-gluon fusion. The
LHC will be a true top factory, producing hundreds of millions of
top quarks every year. With such a large rate, it becomes possible
to accurately measure the total cross section of the top quark pair
production, which provides a good probe of searching for new physics
(NP). The NP effects can modify the $gt\bar{t}$ coupling via quantum
corrections. The non-SM one-loop corrections to the top quark pair
production at hadron colliders have been studied within the general
two-Higgs-doublet model (2HDM)~\citep{Stange:1993td,Zhou:1996dx,Hollik:1997hm,Kao:1999kj}
and the minimal supersymmetric Standard Model (MSSM)~\citep{Kao:1994rn,Li:1995fj,Yang:1996dma,Kim:1996nza,Li:1996jf,Alam:1996mh,Sullivan:1996ry,Li:1997ae,Zhou:1997fw,Hollik:1997fd,Hollik:1997hm,Zhou:1998dc,Kao:1999kj,Berge:2007dz,Ross:2007ez}.
Within these corrections, the Yukawa electroweak radiative correction
is especially interesting because of the existence of the large enhancement
to the Yukawa couplings in the 2HDM~\citep{thdm} and MSSM~\citep{Nilles:1983ge,Haber:1984rc}.
Significant effects indeed were found on both total cross section
and differential cross section distributions, as compared to the one-loop
electroweak corrections in the SM~\citep{Beenakker:1993yr,Kao:1997bs,Stange:1993td,Kuhn:2005it,Bernreuther:2005is,Moretti:2006nf,Kuhn:2006vh,Bernreuther:2006vg}.
In this study we shall examine the leading electroweak corrections
to the top quark pair production in the Littlest Higgs model with
T-parity\ \citep{Cheng:2003ju,Cheng:2004yc,Low:2004xc}.

In Little Higgs models~\citep{ArkaniHamed:2001nc,ArkaniHamed:2002qx,ArkaniHamed:2002qy,Schmaltz:2005ky,Perelstein:2005ka},
the electroweak symmetry is collectively broken and a weak scale Higgs
boson mass is radiatively generated. At one-loop order, the large
quadratically divergent correction to the Higgs boson mass squared
induced by the top quark ($t$) is canceled by its fermionic partner,
and that induced by the electroweak gauge bosons are canceled by their
bosonic partners. Constraints from the low energy precision data,
especially the $\rho$-parameter measurement, require that the symmetry
breaking scale of the Little Higgs models has to be so high that the
predicted phenomenology has little relevance to the current high energy
collider physics program~\citep{Csaki:2002qg,Csaki:2003si,Kilian:2003xt}.
To alleviate the constraints from low energy data, a discrete symmetry,
called T-parity~\citep{Cheng:2003ju,Cheng:2004yc,Low:2004xc}, is
introduced and warrants the $\rho$-parameter to be one at tree-level.
In order to incorporate the T-parity systematically, extra fermion
fields have to be introduced. As a result, we have two sets of particles:
the usual SM particles and an additional $T_{+}$ quark are {}``even''
under the T-parity while the other heavy new particles are {}``odd''.
The SM gauge bosons do not mix with the heavy gauge bosons due to
the T-parity, and the corrections to the low energy observables are
loop-suppressed, consequently, the new particle mass scale $f$ of
the model as low as $500\,{\rm GeV}$ is still allowed~\citep{Hubisz:2005tx}.
Thus the masses of the new particles are at the order of TeV, and
they may cause large quantum corrections to the top quark pair production
at high energy colliders. In this paper, we calculate the \emph{leading}
electroweak (EW) radiative corrections to the anomalous $gt\bar{t}$
couplings by applying the Goldstone-boson equivalence theorem (ET)\ \citep{Cornwall:1974km,Vayonakis:1976vz,Lee:1977eg,Chanowitz:1985hj,Gounaris:1986cr,Yao:1988aj,Bagger:1989fc,Veltman:1989ud,He:1992nga,He:1993yd,He:1993qa,Dobado:1993dg,Dobado:1994vr,He:1994br,Barbieri:1992nz,Barbieri:1992dq}.
We also examine their effects in the $q\bar{q}\to g\to t\bar{t}$
processes at the LHC. The one-loop leading EW corrections to the anomalous
$gt\bar{t}$ coupling are given in terms of the Passarino-Veltman
scalar functions\ \citep{Passarino:1978jh}, which are evaluated
using the library \textsc{looptools (ff})\ \citep{Hahn:1998yk,vanOldenborgh:1989wn,Oldenborgh:1990yc}.

\section{Littlest Higgs model with T-parity}

The Littlest Higgs model with T-parity (LHT) is based on a $SU(5)/SO(5)$
nonlinear sigma model whose low energy Lagrangian is described in
detail in Refs.~\citep{Cheng:2003ju,Cheng:2004yc,Low:2004xc,Hubisz:2004ft,Chen:2006cs}.
With the global symmetry $SU(5)$ being broken down to $SO(5)$ by
a $5\times5$ symmetric tensor at the scale $f$, the gauged $[SU(2)\times U(1)]_{1}\times[SU(2)\times U(1)]_{2}$,
a subgroup of $SU(5)$, is broken to the diagonal $SU(2)_{W}\times U(1)_{Y}$,
a subgroup of $SO(5)$. Four new (T-odd) heavy gauge bosons appear
after the symmetry breaking: the photon partner ($A_{H}$), the $Z$-boson
partner ($Z_{H}$) and the $W^{\pm}$-boson partner ($W_{H}^{\pm}$).
We shall apply the ET to calculate the leading electroweak Yukawa
contributions and adopt the following notations: $h$ is the Higgs
boson; $\pi^{0}(\pi^{\pm})$ is the Goldstone-boson (GB) eaten by
the $Z$-boson ($W$-boson); $\omega^{0}$($\omega^{\pm}$,$\eta$)
is the Goldstone-boson eaten by $Z_{H}$($W_{H}$, $A_{H}$)\ %
\footnote{There is an order of $v^{2}/f^{2}$ mixing between $\omega^{\pm}$
and the $SU(2)$ triplet T-odd scalars $\phi^{\pm}$~\citep{Hubisz:2005tx},
which is neglected in our calculation. %
}. Furthermore, a copy of leptons and quarks with T-odd quantum numbers
are added in order to preserve the T-parity. The T-odd heavy quarks
which contribute to the $gt\bar{t}$ coupling are $t_{-}$, $b_{-}$
and $T_{-}$, which are T-parity partners of the SM top, bottom quarks
and heavy T-even $T_{+}$ quark, respectively. The interactions between
the SM top quark, the $T_{+}$ quark, scalars (the Higgs boson and
GBs), and T-odd quarks could be found by expanding the following effective
Lagrangian,\begin{equation}
{\cal L}_{t}=-\frac{\lambda_{1}}{2\sqrt{2}}f\epsilon_{ijk}\epsilon_{xy}\left[(\bar{Q}_{1})_{i}\Sigma_{jx}\Sigma_{ky}-(\bar{Q}_{2}\Sigma_{0})_{i}\tilde{\Sigma}_{jx}\tilde{\Sigma}_{ky}\right]u_{R}-\lambda_{2}f\left(\bar{U}_{1}U_{R1}+\bar{U}_{2}U_{R2}\right)+h.c.\;\label{eq:Lt}\end{equation}
 and\begin{equation}
{\cal L}_{\kappa}=-\kappa f\left[\bar{\Psi}_{2}\xi\Psi_{c}+\bar{\Psi}_{1}\Sigma_{0}(\Omega\xi^{\dagger}\Omega)\Psi_{c}\right]+h.c.,\,\label{eq:Lk}\end{equation}
 where $\epsilon_{ijk}$ and $\epsilon_{xy}$ are antisymmetric tensors,
and $i,\, j,\, k$ run over $1-3$ and $x,\, y$ over $4-5$; $Q_{1}=(q_{1},U_{1},0,0)^{T}$
, $Q_{2}=(0,0,U_{2},q_{2})^{T}$ where $q_{i}=-\sigma_{2}(u_{i},\, d_{i})^{T}=(id_{i},\,-iu_{i})^{T}$
with $i=1,\,2$; $\Psi_{i}=(q_{i},0,0,0)^{T}$ and $\Psi_{c}=(q_{c},\,\chi_{c},\,\tilde{q}_{c})^{T}$.
(Here, the superscript $T$ denotes taking transpose.) Also, $\Sigma=\xi^{2}\Sigma_{0}$
and $\tilde{\Sigma}=\Sigma_{0}\Omega\Sigma^{\dagger}\Omega\Sigma_{0}$
which is the T-parity transformation of $\Sigma$, where $\xi=exp\{ i\Pi^{a}X^{a}/f\}$,
$X^{a}$ are the broken generators, $\Pi^{a}$ contain the Higgs boson
and all the other GB fields, and \begin{equation}
\Sigma_{0}=\left[\begin{array}{ccc}
0_{2\times2} & 0_{_{2\times1}} & 1_{2\times2}\\
0_{1\times2} & 1 & 0_{1\times2}\\
1_{2\times2} & 0_{2\times1} & 0_{2\times2}\end{array}\right]\,{\rm and}\quad\Omega=\left[\begin{array}{ccc}
1_{2\times2}\\
 & -1\\
 &  & 1_{2\times2}\end{array}\right]_{5\times5}.\label{eq:sigma0}\end{equation}

\begin{table}

\caption{The relevant couplings of the SM top quark and new particles.~\label{tab:heavy}}

\begin{tabular}{c|c|c|c|c|c|c}
\hline 
&
$h-t-T_{+}$ &
$\pi^{0}-t-T_{+}$ &
$\omega^{0}-t-t_{-}$ &
$\eta-t-t_{-}$ &
$\omega^{-}-t-b_{-}$ &
$\eta-t-T_{-}$\tabularnewline
\hline 
$g_{V}$ &
$-\frac{\lambda_{1}^{2}}{2\sqrt{\lambda_{1}^{2}+\lambda_{2}^{2}}}$ &
$i\frac{\lambda_{1}^{2}}{2\sqrt{\lambda_{1}^{2}+\lambda_{2}^{2}}}$ &
$i\frac{\sqrt{2}}{4}\kappa$ &
$-i\frac{\sqrt{10}}{20}\kappa$ &
$i\frac{1}{2}\kappa$ &
$-i\frac{\sqrt{5}}{5}\frac{\lambda_{1}\lambda_{2}}{\sqrt{\lambda_{1}^{2}+\lambda_{2}^{2}}}$\tabularnewline
$g_{A}$ &
$-\frac{\lambda_{1}^{2}}{2\sqrt{\lambda_{1}^{2}+\lambda_{2}^{2}}}$ &
$i\frac{\lambda_{1}^{2}}{2\sqrt{\lambda_{1}^{2}+\lambda_{2}^{2}}}$ &
$i\frac{\sqrt{2}}{4}\kappa$ &
$-i\frac{\sqrt{10}}{20}\kappa$ &
$i\frac{1}{2}\kappa$ &
$i\frac{\sqrt{5}}{5}\frac{\lambda_{1}\lambda_{2}}{\sqrt{\lambda_{1}^{2}+\lambda_{2}^{2}}}$\tabularnewline
\hline
\end{tabular}
\end{table}

For more details of the LHT model, see Refs.~\citep{Cheng:2003ju,Cheng:2004yc,Low:2004xc,Hubisz:2004ft,Chen:2006cs}.
Here, we only list the couplings of the SM top quark and new heavy
particles, which contribute to the loop corrections to the $gt\bar{t}$
coupling, as shown in Table\ \ref{tab:heavy}%
\footnote{Our Feynman rules coincide with the results in Refs.~\citep{Hubisz:2004ft,Belyaev:2006jh,Blanke:2006eb},
up to the ${\cal O}(v/f)$ accuracy.%
}. The coupling of the $\bar{t}FS$ interaction relevant to our calculations
is given as $i(g_{V}+g_{A}\gamma_{5})$, where $F$ ($S$) denotes
the heavy fermion (scalar). There also exist couplings between T-odd
$SU(2)$ triplet scalars $\phi$ to the top quark, but they are neglected
in this work since they are at the ${\cal O}(v/f)$. Since we perform
our calculations in the 't Hooft-Feynman gauge, the mass of the would-be
GB is the same as its corresponding gauge boson. The masses of the
heavy particles are given as follows: \begin{eqnarray}
 &  & m_{t}\sim\frac{\lambda_{1}\lambda_{2}}{\sqrt{\lambda_{1}^{2}+\lambda_{2}^{2}}}v,\,\, m_{T_{+}}\sim\sqrt{\lambda_{1}^{2}+\lambda_{2}^{2}}f,\,\, m_{T-}=\lambda_{2}f,\label{eq:mass1}\\
 &  & m_{\omega^{\pm,0}}\sim gf,\,\, m_{\eta}\sim\frac{g^{\prime}f}{\sqrt{5}},\,\, m_{t-}\simeq m_{b-}\sim\sqrt{2}\kappa f,\label{eq:mass2}\end{eqnarray}
 where $g$ ($g^{\prime}$) is the weak (hypercharge) gauge coupling
strength, and $v\simeq246$ GeV. With those couplings and masses of
the new particles, we now calculate the one-loop corrections to the
$gt\bar{t}$ coupling in the LHT model.

\section{Form factor of $gt\bar{t}$ and one-loop EW corrections in the LHT}

Following the parametrization in Ref.~\citep{Stange:1993td}, the
effective matrix element of $gt\bar{t}$, including the one-loop corrections,
can be written as\begin{equation}
-ig_{s}T^{a}\bar{u}_{t}\Gamma^{\mu}v_{\bar{t}},\label{eq:GenForm0}\end{equation}
 with\begin{equation}
\Gamma^{\mu}=(1+\alpha)\gamma^{\mu}+i\beta\sigma^{\mu\nu}q_{\nu}+\xi\left(\gamma^{\mu}-\frac{2m_{t}}{\hat{s}}q^{\mu}\right)\gamma_{5}.\label{eq:GenForm}\end{equation}
 where the loop-induced form factors $\alpha$, $\beta$ and $\xi$
are usually refereed as the chromo-charge, chromo-magnetic-dipole~%
\footnote{The one-loop non-SM contributions to the $gtt$ chromo-magnetic-dipole
form factor have been recently studied in the literature~\cite{Martinez:2007qf},
where several models are considered, including 2HDM, topcolor assisted
Technicolor model, 331 model and universal extra dimension model.%
} and chromo-anapole, respectively. Here, $g_{s}$ is the strong coupling
strength, $T^{a}$ are the color generators, $q=p_{t}+p_{\bar{t}}$,
and $\hat{s}=q^{2}$. After summing over the final state and averaging
over the initial state colors and spins, the constituent total cross
section of $q\bar{q}\to g\to t\bar{t}$ is~\citep{Stange:1993td}\begin{equation}
\hat{\sigma}=\frac{8\pi\alpha_{s}^{2}}{27\hat{s}^{2}}\sqrt{1-\frac{4m_{t}^{2}}{\hat{s}}}\,\biggl\{\hat{s}+2m_{t}^{2}+2\Re\left[(\hat{s}+2m_{t}^{2})\alpha+3m_{t}\hat{s}\,\beta\right]\biggr\},\,\label{eq:Hardxsect}\end{equation}
 where $\alpha_{s}\equiv g_{s}^{2}/(4\pi)$, and $\Re$ denotes taking
its real part. Note that $\xi$ does not contribute to the total cross
section at this order, as a result of the interference with the Born
matrix element, but for completeness we will present the analytical
expressions of those three form factors in the LHT model below.

\begin{figure}[b]
 \includegraphics[clip,scale=0.5]{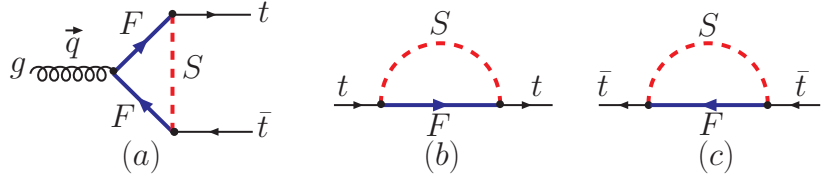}

\caption{Feynman diagrams of the one-Loop corrections to the $gt\bar{t}$
coupling in the LHT model~\label{fig:feynman-gtt}.}
\end{figure}

At the one-loop level, the $gt\bar{t}$ coupling receives two kinds
of quantum corrections: one is the irreducible triangle-loop correction
(Fig.\ \ref{fig:feynman-gtt}a), another is the self-energy correction
to the external top quark lines (Figs.\ \ref{fig:feynman-gtt}b\ and\ \ref{fig:feynman-gtt}c).
For simplicity, we use the particles running inside the loop to represent
the corresponding loop correction diagram. For example, Fig.\ \ref{fig:feynman-gtt}(a)
are denoted as $\left(F,F,S\right)$. In the LHT model, the diagrams
contributing to the anomalous $gt\bar{t}$ coupling are given by $\left(T_{+},T_{+},h/\pi^{0}\right)$,
$\left(t_{-},t_{-},\eta/\omega^{0}\right)$, $\left(b_{-},b_{-},\omega^{\pm}\right)$
and $\left(T_{-},T_{-},\eta\right)$. The coupling strength of the
$gF\bar{F}$ vertex is just the usual strong coupling while the $\bar{t}FS$
couplings are explicitly given in Table\ \ref{tab:heavy}.

We use the dimensional regularization scheme to regulate the ultraviolet
divergences and adopt the on-mass-shell renormalization scheme to
renormalize the electroweak parameters. In this scheme, the wave function
renormalization corrections of the external top quark legs are canceled
by the corresponding counterterms. We will regularize the ultraviolet
divergences in our calculation by dimensional regularization with
the regulator defined by $\Delta=\frac{1}{\epsilon}-\gamma_{E}+\ln4\pi$,
where $2\epsilon\equiv4-n$, $n$ is the dimension of the space-time
and $\gamma_{E}$ is the Euler constant. As we are calculating the
leading EW corrections to the $gt\bar{t}$ coupling, we do not need
to introduce the counterterm for the strong coupling. By introducing
appropriate counterterms, one can easily deduce the renormalized vertex
of $gt\bar{t}$ as \begin{equation}
-ig_{s}T^{a}\bar{u}_{t}\left(\gamma^{\mu}+\delta\Gamma_{{\rm ren}}^{\mu}\right)v_{t},\end{equation}
 where \begin{equation}
\delta\Gamma_{ren}^{\mu}=\gamma^{\mu}\left(\delta Z_{V}^{t}+\delta Z_{A}^{t}\gamma_{5}\right)+\delta\Gamma_{\triangle}^{\mu}\label{eq:efc_corrections}\end{equation}
 Here, $\delta Z_{V,A}^{t}$ denote the wave function renormalization
constants of the external top quark lines, which are defined by \[
Z_{t}\equiv1+\delta Z_{t}=1+\delta Z_{V}^{t}+\delta Z_{A}^{t}\gamma_{5},\]
 while $\delta\Gamma_{\triangle}$ denotes the triangle loop corrections
to the vertex. Clearly, the $\delta Z_{V}$ counter terms only contribute
to the form factor $\alpha$, the $\delta Z_{A}$ counter terms only
contribute to the form factor $\xi$, but the vertex corrections $\delta\Gamma_{\triangle}$
contribute to all three form factors. We thus write the form factors
as follows,\begin{equation}
\alpha=\alpha_{\triangle}+\delta Z_{V},\qquad\beta=\beta_{\triangle},\qquad\xi=\xi_{\triangle}+\delta Z_{A},\end{equation}
 where $\alpha_{\triangle}$, $\beta_{\triangle}$ and $\xi_{\triangle}$
denote the coefficients of the $\gamma^{\mu}$, $\sigma^{\mu\nu}q_{\nu}$
and $\gamma^{\mu}\gamma_{5}$ terms in $\delta\Gamma_{\triangle}^{\mu}$,
respectively. Note that there is an additional term $q^{\mu}\gamma_{5}$
in $\delta\Gamma_{\triangle}^{\mu}$. After adding the $\delta Z_{A}$
counter terms, we can write the combination of $\gamma^{\mu}\gamma_{5}$
and $q^{\mu}\gamma_{5}$ in a compact form as the $\xi$ term in Eq.\ (\ref{eq:GenForm}),
which is guaranteed by the Ward identity for the conservation of QCD
current.

Consider the renormalization constants. The wave function renormalization
constants can be determined from the top quark self-energy diagrams,
cf. Figs.\ \ref{fig:feynman-gtt}(b,\ c), which can be decomposed
as follows:\begin{equation}
\Sigma\left(\pslash{p}\right)=\pslash{p}\left[\Sigma_{V}\left(p^{2}\right)+\Sigma_{A}\left(p^{2}\right)\gamma_{5}\right]+m_{t}\Sigma_{S}\left(p^{2}\right).\end{equation}
 In the on-shell scheme, the finite parts of the counter terms are
determined by the requirement that the residue of the fermion propagator
is equal to one, which fixes the wave function renormalization constraints
by\begin{eqnarray}
\delta Z^{V} & = & -\Sigma_{V}\left(p^{2}=m_{t}^{2}\right)-2m_{t}^{2}\frac{\partial}{\partial p^{2}}\left.\left(\Sigma_{V}+\Sigma_{S}\right)\right|_{p^{2}=m_{t}^{2}},\\
\delta Z^{A} & = & -\Sigma_{A}\left(p^{2}=m_{t}^{2}\right).\end{eqnarray}
 In the LHT model, they are given by\begin{eqnarray}
\delta Z^{V} & = & \frac{1}{16\pi^{2}}\frac{g_{V}^{2}+g_{A}^{2}}{2m_{t}^{2}}\left\{ A_{0}\left(m_{S}^{2}\right)-A_{0}\left(m_{F}^{2}\right)+\left(m_{F}^{2}-m_{S}^{2}-m_{t}^{2}\right)B_{0}\left(m_{t}^{2}\right)\right\} \nonumber \\
 & + & \frac{1}{16\pi^{2}}\biggl[\left(g_{V}^{2}+g_{A}^{2}\right)\left(-m_{t}^{2}+m_{S}^{2}-m_{F}^{2}\right)-\left(g_{V}^{2}-g_{A}^{2}\right)2m_{t}m_{F}\biggr]B_{0}^{\prime}\left(m_{t}^{2}\right),\\
\delta Z^{A} & = & \frac{1}{16\pi^{2}}\frac{g_{V}g_{A}}{m_{t}^{2}}\left\{ A_{0}\left(m_{S}^{2}\right)-A_{0}\left(m_{F}^{2}\right)+\left(m_{F}^{2}-m_{S}^{2}+m_{t}^{2}\right)B_{0}\left(m_{t}^{2}\right)\right\} ,\end{eqnarray}
 where $A_{0}$ and $B_{0}$ are the well-known one-point and two-point
scalar functions\ \citep{Passarino:1978jh}. We also introduce the
following shorthand notations,\begin{equation}
B_{0}\left(m_{t}^{2}\right)\equiv B_{0}\left(m_{t}^{2};m_{S}^{2},m_{F}^{2}\right),\qquad B_{0}^{\prime}\left(m_{t}^{2}\right)\equiv\frac{\partial}{\partial p^{2}}\left.B_{0}\left(p^{2};m_{S}^{2},m_{F}^{2}\right)\right|_{p^{2}=m_{t}^{2}}.\end{equation}
 where $m_{S}$ ($m_{F}$) is the mass of the scalar (fermion) in
the loop.

Now considering the vertex corrections $\delta\Gamma_{\triangle}^{\mu}$,
which we decompose into the form factors $\alpha_{\triangle}$, $\beta_{\triangle}$
and $\xi_{\triangle}$, as listed below. The form factor $\alpha_{\triangle}$
is given by\begin{eqnarray}
\alpha_{\triangle} & = & -\frac{g_{V}g_{V}^{*}}{16\pi^{2}}\left\{ \alpha_{1}+\alpha_{2}B_{0}\left(\hat{s}\right)+\alpha_{3}B_{0}\left(m_{t}^{2}\right)+\alpha_{4}C_{0}\right\} \nonumber \\
 &  & -\frac{g_{A}g_{A}^{*}}{16\pi^{2}}\left\{ \alpha_{1}^{\prime}+\alpha_{2}^{\prime}B_{0}\left(\hat{s}\right)+\alpha_{3}^{\prime}B_{0}\left(m_{t}^{2}\right)+\alpha_{4}^{\prime}C_{0}\right\} ,\end{eqnarray}
 where\begin{eqnarray}
\alpha_{1} & = & \frac{\hat{s}}{2\left(\hat{s}-4m_{t}^{2}\right)}+\frac{2}{\hat{s}-4m_{t}^{2}}\left[-A_{0}(m_{S}^{2})+A_{0}(m_{F}^{2})\right],\\
\alpha_{2} & = & \frac{1}{2(\hat{s}-4m_{t}^{2})^{2}}\biggl[-16m_{t}^{4}-32m_{F}m_{t}^{3}+(-16m_{F}^{2}+16m_{S}^{2}+14\hat{s})m_{t}^{2}\nonumber \\
 &  & \qquad\qquad\qquad+8m_{F}\hat{s}m_{t}-\hat{s}^{2}-2m_{F}^{2}\hat{s}+2m_{S}^{2}\hat{s}\biggr],\\
\alpha_{3} & = & \frac{1}{2(\hat{s}-4m_{t}^{2})^{2}}\biggl[32m_{F}m_{t}^{3}+(32m_{F}^{2}-32m_{S}^{2}-6\hat{s})m_{t}^{2}\nonumber \\
 &  & \qquad\qquad\qquad-8m_{F}\hat{s}m_{t}-2\hat{s}(m_{F}^{2}-m_{S}^{2})\biggr],\\
\alpha_{4} & = & \frac{1}{2(\hat{s}-4m_{t}^{2})^{2}}\biggl[16m_{t}^{6}+32m_{F}m_{t}^{5}+(32m_{F}^{2}-32m_{S}^{2}-6\hat{s})m_{t}^{4}\nonumber \\
 &  & +(32m_{F}^{3}-32m_{F}m_{S}^{2}-24m_{F}\hat{s})m_{t}^{3}\nonumber \\
 &  & +(16m_{F}^{4}+16m_{S}^{4}-32m_{F}^{2}m_{S}^{2}+2\hat{s}^{2}-28m_{F}^{2}\hat{s}+20m_{S}^{2}\hat{s})m_{t}^{2}\nonumber \\
 &  & +(4m_{F}\hat{s}^{2}-8m_{F}^{3}\hat{s}+8m_{F}m_{S}^{2}\hat{s})m_{t}+2m_{F}^{2}\hat{s}^{2}+2m_{F}^{4}\hat{s}+2m_{S}^{4}\hat{s}-4m_{F}^{2}m_{S}^{2}\hat{s}\biggr],\end{eqnarray}
 and \begin{equation}
\alpha_{1}^{\prime}=\alpha_{1},\qquad\alpha_{2,3,4}^{\prime}=\alpha_{2,3,4}\biggr|_{m_{F}\to-m_{F}}.\end{equation}
 Here we introduce the following shorthand notations,\begin{equation}
B_{0}\left(\hat{s}\right)\equiv B_{0}\left(\hat{s};m_{t}^{2},m_{t}^{2}\right),\qquad C_{0}\equiv C_{0}\left(m_{t}^{2},\hat{s};m_{S}^{2},m_{F}^{2},m_{F}^{2}\right),\end{equation}
 where $C_{0}\left(...\right)$ is the usual three-point scalar function\ \citep{Passarino:1978jh}.
The form factor $\beta_{\triangle}$ is given by\begin{eqnarray}
\beta_{\triangle} & = & \frac{g_{V}g_{V}^{*}}{16\pi^{2}}\left\{ \beta_{1}+\beta_{2}B_{0}\left(\hat{s}\right)+\beta_{3}B_{0}\left(m_{t}^{2}\right)+\beta_{4}C_{0}\right\} \nonumber \\
 & + & \frac{g_{A}g_{A}^{*}}{16\pi^{2}}\left\{ \beta_{1}^{\prime}+\beta_{2}^{\prime}B_{0}\left(\hat{s}\right)+\beta_{3}^{\prime}B_{0}\left(m_{t}^{2}\right)+\beta_{4}^{\prime}C_{0}\right\} ,\end{eqnarray}
 where \begin{eqnarray}
\beta_{1} & = & \frac{m_{t}}{\hat{s}-4m_{t}^{2}}+\frac{1}{m_{t}(\hat{s}-4m_{t}^{2})}\left[-A_{0}(m_{S}^{2})+A_{0}(m_{F}^{2})\right],\\
\beta_{2} & = & \frac{1}{(\hat{s}-4m_{t}^{2})^{2}}\left[2m_{t}^{3}-8m_{F}m_{t}^{2}+(-6m_{F}^{2}+6m_{S}^{2}+\hat{s})m_{t}+2m_{F}\hat{s}\right],\\
\beta_{3} & = & \frac{1}{m_{t}(\hat{s}-4m_{t}^{2})^{2}}\biggl[-2m_{t}^{4}+8m_{F}m_{t}^{3}+(10m_{F}^{2}-10m_{S}^{2}-\hat{s})m_{t}^{2}\nonumber \\
 &  & -2m_{F}\hat{s}m_{t}+(m_{S}^{2}-m_{F}^{2})\hat{s}\biggr],\\
\beta_{4} & = & \frac{-2}{(\hat{s}-4m_{t}^{2})^{2}}\biggl[m_{t}^{5}+4m_{F}m_{t}^{4}+(2m_{F}^{2}+2m_{S}^{2}-\hat{s})m_{t}^{3}-m_{F}(4m_{F}^{2}-4m_{S}^{2}+\hat{s})m_{t}^{2}\nonumber \\
 &  & +\left(-3m_{F}^{4}+m_{F}^{2}(6m_{S}^{2}+\hat{s})-3m_{S}^{4}-2m_{S}^{2}\hat{s}\right)m_{t}+m_{F}(m_{F}^{2}-m_{S}^{2})\hat{s}\biggr],\end{eqnarray}
 and\begin{equation}
\beta_{1}^{\prime}=\beta_{1},\qquad\beta_{2,3,4}^{\prime}=\beta_{2,3,4}\biggr|_{m_{F}\to-m_{F}}.\end{equation}
 Finally, the form factor $\xi_{\triangle}$ is given by\begin{equation}
\xi_{\triangle}=-\frac{g_{V}g_{A}^{*}}{16\pi^{2}}\left\{ -1+\xi_{1}B_{0}\left(\hat{s}\right)+\xi_{2}B_{0}\left(m_{t}^{2}\right)+\xi_{3}C_{0}\right\} ,\end{equation}
 where\begin{eqnarray}
\xi_{1} & = & \frac{1}{\hat{s}-4m_{t}^{2}}\left[2m_{t}^{2}-2m_{S}^{2}+2m_{F}^{2}+\hat{s}\right],\\
\xi_{2} & = & \frac{-2}{\hat{s}-4m_{t}^{2}}\left[m_{F}^{2}-m_{S}^{2}+3m_{t}^{2}\right],\\
\xi_{3} & = & \frac{-2}{\hat{s}-4m_{t}^{2}}\left[m_{t}^{4}-(2m_{F}^{2}+2m_{S}^{2}+\hat{s})m_{t}^{2}+m_{S}^{4}+m_{F}^{4}-2m_{F}^{2}m_{S}^{2}+m_{F}^{2}\hat{s}\right].\end{eqnarray}

\section{Numerical results~\label{sec:Numerical}}

The model parameters for the numerical evaluation are $\lambda_{1}$,
$\lambda_{2}$, $\kappa$ and $f$ . As $\lambda_{1}$ and $\lambda_{2}$
are related by the mass of the top quark, cf. Eq.\ (\ref{eq:mass1}),
we could choose either one as the input parameter, and in this study
$\lambda_{1}$ is chosen. As pointed out from the partial wave study
in Ref.~\citep{Belyaev:2006jh}, $\lambda_{1}$ should be bounded
in the region $0.71\lesssim\lambda_{1}\lesssim2.51$. Furthermore,
if $\kappa$ is not universal for quark and lepton sectors, as studied
in Ref.~\citep{Cao:2007pv}, the upper bound for $\kappa$ of the
quark sector from the constrains of four-fermion operators could be
quite loose even for a low $f$ value, say $f\sim500$ GeV. For illustration,
we choose the values of the parameters as follows:\begin{eqnarray*}
 &  & \lambda_{1}=2.5,\,\,\,\,\kappa=5,\,\,\,\, f=500\,{\rm GeV},\,\,\,\, m_{t}=175\,{\rm GeV},\\
 &  & m_{W}=80.4\,{\rm GeV},\, m_{Z}=91.2\,{\rm GeV},\, m_{h}=120(500)\,{\rm GeV},\end{eqnarray*}
 where $m_{W}$, $m_{Z}$ and $m_{h}$ denote the masses of the $W$-boson,
$Z$-boson and Higgs boson, respectively, and the bottom quark is
considered as massless throughout this work. With the chosen parameters,
the masses of new heavy particles are given by\begin{eqnarray*}
 &  & m_{T_{+}}=1302\,{\rm GeV},\quad m_{T_{-}}=364\,{\rm GeV},\\
 &  & m_{t_{-}}\simeq m_{b_{-}}=3536\,{\rm GeV},\quad m_{\omega^{\pm,\,0}}=327\,{\rm GeV},\quad m_{\eta}=78\,{\rm GeV}.\end{eqnarray*}

\begin{figure}
\includegraphics[clip,scale=0.6]{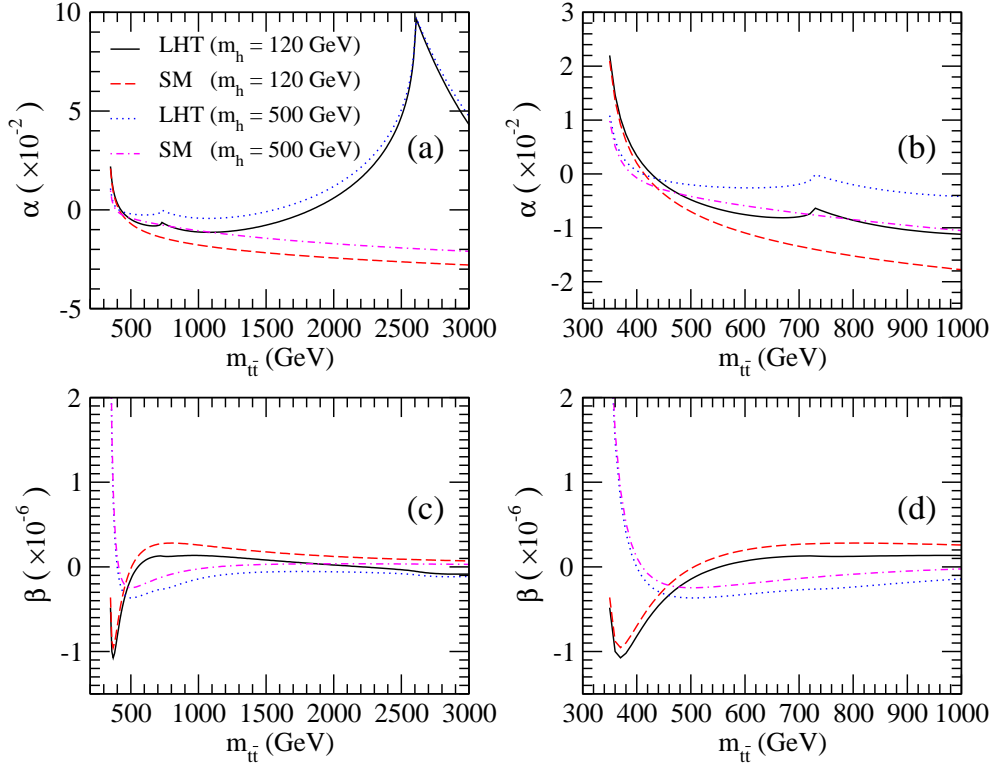}

\caption{Dependence of the invariant mass of the top quark pair in form factors
in both the LHT model and the SM: (a) and (b) $\alpha$; (c) and (d)
$\beta$. (b) and (d) is the same as (a) and (c), respectively, but
focusing on the small $m_{t\bar{t}}$ region. \ \label{fig:alpha-beta}}
\end{figure}

Since, as a result of the interference with the Born matrix element,
$\xi$ does not contribute, we need only the form factors $\alpha$
and $\beta$, which depend on both the couplings ($g_{V}$ and $g_{A}$)
and the masses of the scalars and fermions flowing in the loops. We
split the form factors in the LHT, $\alpha_{LHT}$ and $\beta_{LHT}$,
as follows:\begin{equation}
\alpha_{LHT}=\alpha_{SM}+\alpha_{HEAVY},\qquad\beta_{LHT}=\beta_{SM}+\beta_{HEAVY},\end{equation}
 where the subscript \emph{SM} and \emph{HEAVY} denote contributions
to form factors which are induced by the SM loops and the new heavy
particle loops, respectively. In Figs.\ \ref{fig:alpha-beta}(a)
and (c), we present the values of form factors $\alpha$ and $\beta$
as a function of the invariant mass of the top quark pair system,
respectively. In order to investigate the dependence of the SM Higgs
boson mass, we also choose two different Higgs boson masses: $m_{h}=120\,{\rm GeV}$
and $m_{h}=500\,{\rm GeV}$. We note a few interesting points listed
as follows:

\begin{itemize}
\item For $m_{t\bar{t}}>500\,{\rm GeV}$, $\alpha_{SM}$ is negative but
$\alpha_{HEAVY}$ is positive. Furthermore, in the region of $400\,{\rm GeV}<m_{t\bar{t}}<2000\,{\rm GeV}$,
$\alpha_{HEAVY}\simeq\left|\alpha_{SM}\right|$. Therefore, their
sum, $\alpha_{LHT}$, is around zero. The small kink in $\alpha_{HEAVY}$
near $m_{t\bar{t}}\sim2m_{T_{-}}$\,GeV is due to the threshold effect
from producing the $T_{-}{\bar{T}_{-}}$ pair. However, in the large
$m_{t\bar{t}}$ region, e.g. $m_{t\bar{t}}>2500\,{\rm GeV}$, $\alpha_{HEAVY}$
receives a large corrections from the $(T_{+},T_{+},h/\pi^{0})$ loops,
and is much larger than $\left|\alpha_{SM}\right|$. In particular,
$\alpha_{HEAVY}$ reaches its maximum around the threshold region,
i.e. $m_{t\bar{t}}\sim2m_{T_{+}}$. As a result, $\alpha_{LHT}$ is
positive and much larger than $\alpha_{SM}$ in the large $m_{t\bar{t}}$
region, see the (black) solid line ($m_{h}=120\,{\rm GeV}$) and the
(blue) dotted line ($m_{h}=500\,{\rm GeV}$) in Fig.~\ref{fig:alpha-beta}
(a). In the small $m_{t\bar{t}}$ region, i.e. $m_{t\bar{t}}<500\,{\rm GeV}$,
$\alpha_{HEAVY}$ is negligible and $\alpha_{LHT}\simeq\alpha_{SM}$. 
\item The form factor $\beta_{HEAVY}$ is always negative, see the (black)
solid line (LHT) and the (red) dashed line (SM) in Fig.\ \ref{fig:alpha-beta}(d).
In the large $m_{t\bar{t}}$ region, both $\beta_{LHT}$ and $\beta_{SM}$
are negligible. Note that the chromo-magnetic-dipole form factor $\beta$
can contribute to the branching ratio of $b\to s\gamma$ process~\citep{Hewett:1993em,Martinez:1996cy,Martinez:2001qs},
and our numerical results are consistent with the current bounds~\citep{Martinez:2001qs}. 
\end{itemize}
\begin{figure}
\includegraphics[clip,scale=0.5]{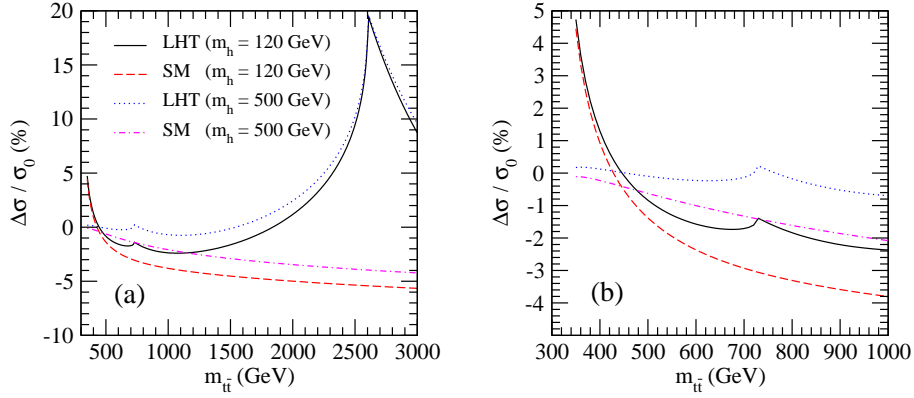}

\caption{The ratio of the one-loop leading EW correction to the Born level
total cross section of $q\bar{q}\to g\to t\bar{t}$ at the LHC. (b)
is the same as (a) but focusing on the small $m_{t\bar{t}}$ region.\ \label{fig:R_sig} }
\end{figure}

Below, we will examine the effects of the leading EW corrections on
the top quark pair production at the LHC. For that, we calculate the
differential cross section, $d\sigma/dm_{t\bar{t}}$, given by \[
\frac{d\sigma}{dm_{t\bar{t}}}=\int dx_{1}dx_{2}\left\{ f_{q/p}\left(x_{1},Q\right)f_{\bar{q}/p}\left(x_{2},Q\right)\frac{d\hat{\sigma}}{dm_{t\bar{t}}}\left(q\bar{q}\to t\bar{t}\right)+\left(x_{1}\leftrightarrow x_{2}\right)\right\} ,\]
 where $\hat{\sigma}$ labels the hard process cross section, and
$f_{q/p}\left(x,Q\right)$ denotes the parton distribution function
of finding the parton $q$ in the colliding proton with the momentum
fraction $x$. $Q$ is the factorization scale of the hard scattering
process. In our calculations, we use the CTEQ 6.1 parton distribution
functions~\citep{Pumplin:2002vw}. We note that at the LHC, the dominant
mechanism for top quark pair production is via gluon-gluon fusion,
i.e., $gg\to t\bar{t}$. Nevertheless, in this work, we focus on the
new physics effect predicted by the LHT to top quark pair production
cross section in the quark and anti-quark scattering processes. To
examine in detail the effect of leading EW corrections, we calculate
the relative corrections defined as \begin{equation}
\frac{\Delta\sigma}{\sigma^{0}}\equiv\left(\frac{d\sigma}{dm_{t\bar{t}}}-\frac{d\sigma_{0}}{dm_{t\bar{t}}}\right)/\frac{d\sigma_{0}}{dm_{t\bar{t}}},\label{eq:relative_correction}\end{equation}
 where $\sigma_{0}$ denotes the tree-level SM cross section. Fig.\ \ref{fig:R_sig}(a)
shows our numerical results, while Fig.\ \ref{fig:R_sig}(b) reveals
the details of the small $m_{t\bar{t}}$ region of Fig.\ \ref{fig:R_sig}(a).
It is clear that the relative corrections are dominated by $\alpha$,
because $\alpha$ is much larger than $\beta$. Again, we find that
the negative EW corrections in the SM are almost canceled by the positive
EW corrections from the new heavy particle loops in the LHT model
in the region of $m_{t\bar{t}}<2000\,{\rm GeV}$. In the large $m_{t\bar{t}}$
region, the leading EW corrections in the LHT model could increase
the cross section by about $20\%$. However, such a deviation might
hardly be recognized as the cross section drops rapidly with increasing
$m_{t\bar{t}}$. Moreover, bearing in mind that the top quark pair
production at the LHC is predominately via the gluon-gluon fusion
process, a systematic study including the $gg\to t\bar{t}$ process
is in order and will be presented in the forthcoming paper.

\section{Conclusion}

In this paper, we calculate the leading electroweak (EW) corrections
to the anomalous $gt\bar{t}$ couplings in the LHT model by applying
the Goldstone-boson equivalence theorem, and further examine their
effects on the top quark pair production cross section via quark annihilation
processes at the LHC. We found that the negative EW corrections in
the SM are partially canceled by the positive EW corrections from
the new heavy particle loops in the LHT model. The net one-loop electroweak
correction is close to zero in the range of $500\,{\rm GeV}<m_{t\bar{t}}<2000\,{\rm GeV}$.
For a larger value of $m_{t\bar{t}}$, the new heavy particle loop
correction dominates. A complete study including the electroweak corrections
to the top quark pair production via the gluon-gluon fusion process
will be presented in the forthcoming paper.

\begin{acknowledgments}
Q.-H. Cao is supported in part by the U.~S.~Department of Energy
under Grant No.~DE-FG03-94ER40837. C.-P. Yuan and C.-R. Chen are
supported in part by the U.S. National Science Foundation under Grant
No. PHY-0555545. F. Larios thanks Conacyt for support.

\bigskip{}

\end{acknowledgments}
Note added: While finalizing the write-up of this work, we are aware
of a paper~\citep{Ding:2008nh}, in which the chromo-magnetic-dipole
form factor in the LHT model is also studied.

\bibliographystyle{apsrev} \bibliographystyle{apsrev} \bibliographystyle{apsrev}
\bibliography{reference}

\end{document}